\documentclass[aps,pra,showpacs,twocolumn,preprintnumbers]{revtex4-1}
\usepackage{graphicx,graphics,epsfig,subfigure,times,bm,bbm,amssymb,amsmath,amsfonts,amsthm,mathrsfs,MnSymbol}
\usepackage{dcolumn}
\usepackage{color}
\usepackage{dsfont}
\usepackage{multirow}
\usepackage[colorlinks,bookmarks=true,citecolor=blue,linkcolor=red,urlcolor=blue]{hyperref}
\usepackage{verbatim}

\newcommand{\be}{\begin{equation}}
\newcommand{\ee}{\end{equation}}
\newcommand{\ba}{\begin{eqnarray}}
\newcommand{\ea}{\end{eqnarray}}
\newcommand{\ignore}[1]{}

\begin{document}

\title{Euler characteristic number of the energy band and the reason for its
non-integer values}
\author{Yu-Quan Ma}
\affiliation{School of Science, Beijing Information Science and Technology
	University, Beijing 100192, China}
\date{\today }

\begin{abstract}
The topological Euler characteristic number of the energy band proposed in
our previous work (see Yu-Quan Ma et al., arXiv:1202.2397; EPL 103, 10008
(2013)) has been recently experimentally observed by X. Tan et al., Phys.
Rev. Lett. \textbf{122}, 210401 (2019), in which a topological phase
transition in a time-reversal-symmetric system simulated by the
superconducting circuits is witnessed by the Euler number of the occupied
band instead of the vanishing Chern number. However, we note that there are
some confusions about the non-integer behaviors of the Euler number in the
topological trivial phase. In this paper, we show that the reason is
straightforward because the quantum metric tensor $g_{\mu \nu} $ is actually
positive semi-definite. In a general two-dimensional two-band system, we can
proved that: (1) If the phase is topological trivial, then the quantum
metric must be degenerate (singular)~--- $\det {g_{\mu \nu} }=0$ in some
region of the first Brillouin zone. This leads to the invalidity of the
Gauss-Bonnet formula and exhibits an ill-defined ``non-integer Euler
number''; (2) If the phase is topological nontrivial with a non-vanishing
Berry curvature, then the quantum metric will be a positive definite Riemann
metric in the entire first Brillouin zone. Therefore the Euler number of the
energy band will be guaranteed an even number $\chi=2(1-g)$ by the
Gauss-Bonnet theorem on the closed two-dimensional Bloch energy band
manifold with the genus $g$, which provides an effective topological index
for a class of nontrivial topological phases.
\end{abstract}

\maketitle

\section{Introduction}

In a recent paper~\cite{xtan}, X. Tan et al. report a direct experimental
measurement of the quantum metric tensor in a tunable superconducting
circuits system and characterize a topological phase transition in the
simulated time-reversal-symmetric system by the Euler characteristic number
of the energy band instead of the vanishing Chern number~\cite{Ma2013,xtan}.
However, there are some confusions about the Euler number calculated from
the general formula (see Ref.~\cite{Ma2013}) exhibiting some non-integer
behaviors in the topological trivial phase of a two-band Hamiltonian
discussed in Ref.~\cite{Ma2013,xtan}.

We also note that there are some recent efforts~\cite{zhu note} trying to
attribute the \textquotedblleft non-integer Euler number\textquotedblright\
behaviors to the missing of the boundary contribution of the $\vec{d}(%
\mathbf{k})/d(\mathbf{k})$ surface itself when the Hamiltonian parameter $%
h>1 $. Concretely, they study this problem in the approach of Fubini-Study
metric on the unit sphere $S^{2}$ and find that the surface of $\vec{d}(%
\mathbf{k})/d(\mathbf{k})$ will become unclosed when the parameter $h>1$,
and hence, they believe that the Euler number of the unclosed surface of $%
\vec{d}(\mathbf{k})/d(\mathbf{k})$ should be the corrected Euler number of
the energy band manifold when the parameter $h>1$.

However, we would like to point out that the role of the mapping $\vec{d}(%
\mathbf{k})/d(\mathbf{k})$ unit sphere is actually to serve as a quantum
states space $\mathcal{C}P^{1}\cong S^{2}$ for the Fubini-Study metric, but
not to serve as a parameter manifold for the quantum metric tensor~\cite%
{DCAJ}. Strictly speaking, the Fubini-Study metric is only a special case of
the quantum metric tensor. Unlike the quantum metric on the parameter
manifold, the Fubini-Study metric on the projective Hilbert space $\mathcal{%
P(H)}$ always defines a proper Riemann metric. However, as we show later,
the singularity of the quantum metric is closely related to the zero points
of the Berry curvature. There is no correspondence between the topology of
the $\vec{d}(\mathbf{k})/d(\mathbf{k}) $ surface itself and the
two-dimensional Bloch energy band manifold on the first Brillouin zone.

In the Haldane model, as a new example, we find that the Euler number of the
filled band is $\chi =2$ corresponding to the Chern number $C_{1}=\pm 1$,
and $\chi =$ \textquotedblleft non-integer\textquotedblright\ corresponding
to the $C_{1}=0$; Furthermore, an important fact is that the surfaces of $%
\vec{d}(\mathbf{k})/d(\mathbf{k})$ for the Haldane model Hamiltonian are all
unclosed, no matter in the topological nontrivial phase or in the trivial
phase. Meanwhile, it can be seen directly that, in the topological
nontrivial phase, the Euler number for the unclosed surface $\vec{d}(\mathbf{%
k})/d(\mathbf{k})$ itself must be zero due to existing two boundaries on its
surface, which makes it topological equivalent to the $S^{1}$ ring with the
Euler number $\chi =0$. Obviously, there is no correspondence between the
Euler number of the $\vec{d}(\mathbf{k})/d(\mathbf{k})$ surface itself and
the Euler number of the Bloch band manifold (for details see Sec.~\ref{example b}).

In this paper, we show the real reason is that the quantum metric $g_{\mu
\nu }$, as the real part of the U(1) gauge invariant quantum geometric
tensor, is strictly a positive semi-definite metric. When studying the local
and global properties for the Riemann structure of the quantum states
manifold, i.e., the Bloch band manifold on the first Brillouin zone, we
should confine the quantum metric in a positive definite parameter region,
where it serves as a proper Riemann metric. More concretely, in a general
two-dimensional two-band system, we can proved that:

(1) If the phase is topological trivial, then the quantum metric must be
degenerate (or singular, its inverse does not exist), that is, $\det {g_{\mu
\nu }}=0$, in some region of the first Brillouin zone, where the Berry
curvature vanishes simultaneously. This leads to the invalidity of the
Gauss-Bonnet formula, and not surprisingly, exhibits an ill-defined
\textquotedblleft non-integer Euler number\textquotedblright ;

(2) If the phase is topological nontrivial with a non-vanishing Berry
curvature, then the quantum metric must be positive definite in the entire
first Brillouin zone, and the Euler number of the energy band will be
guaranteed an even number $\chi =2(1-g)$ by the Gauss-Bonnet theorem on the
closed two-dimensional Bloch energy band manifold with the genus $g$, which
provides an effective topological index for a class of nontrivial
topological phases.

\section{Preliminary}

\subsection{Quantum metric $g_{\protect\mu \protect\nu }$ is positive
semi-definite}

The quantum metric $g_{\mu \nu }$ can be defined as the real part of a $U(1)$
gauge invariant quantum geometric tensor $Q_{\mu {\nu }}:=\langle {\partial
_{\mu }\Psi }\left( \mathbf{k}\right) |\left[ \mathbb{I}-\mathcal{P}\left( 
\mathbf{k}\right) \right] \left\vert {\partial _{{\nu }}\Psi }\left( \mathbf{%
k}\right) \right\rangle $, where the projection operator $\mathcal{P}\left( 
\boldsymbol{k}\right) :=\left\vert {\Psi }\left( \mathbf{k}\right)
\right\rangle \left\langle {\Psi }\left( \mathbf{k}\right) \right\vert $,
and $\mu \left( \nu \right) $ denotes the component $k^{\mu \left( \nu
\right) }$ in the parameter manifold $\mathcal{M}_{\mathbf{k}}$. It is
readily to be verified that $g_{\mu \nu }=\Re \langle {\partial _{\mu }\Psi }%
\left( \mathbf{k}\right) |\left[ \mathbb{I}-\mathcal{P}\left( \mathbf{k}%
\right) \right] \left\vert {\partial _{{\nu }}\Psi }\left( \mathbf{k}\right)
\right\rangle ={\mbox{Tr\,}}\left[ {\partial _{\mu }}\mathcal{P}\left( 
\mathbf{k}\right) {\partial _{{\nu }}}\mathcal{P}\left( \mathbf{k}\right) %
\right] $. For a nonzero vector $V\left( \mathbf{k}\right) \in T_{\mathcal{M}%
}$, the norm of $V\left( \mathbf{k}\right) $ measured by the $g_{\mu \nu }$
is given by~\cite{Zanardi} 
\begin{eqnarray}
\left\Vert V\left( \mathbf{k}\right) \right\Vert ^{2} &=&g_{\mu \nu }V^{\mu
}V^{\nu }={\mbox{Tr\,}}\left[ \left( V^{\mu }{\partial _{\mu }}\mathcal{P}%
\right) \left( V^{\nu }{\partial _{{\nu }}}\mathcal{P}\right) \right]  \notag
\\
&=&\sum_{mn}\langle {n}|V^{\mu }{\partial _{\mu }}\mathcal{P}\left\vert {m}%
\right\rangle \cdot \langle {n}|V^{\nu }{\partial _{{\nu }}}\mathcal{P}%
\left\vert {m}\right\rangle ^{\ast }\geqslant 0.  \label{metric}
\end{eqnarray}%
Clearly, it is possible that a nonzero vector $V\left( \mathbf{k}\right) $
can be endowed with a zero norm by the metric. Hence the quantum metric $%
g_{\mu \nu }$ is actually positive semi-definite.

\subsection{Cauchy--Schwartz inequality for the determinant of quantum
geometric tensor}

\emph{Theorem 1.}---\ For a two-dimensional $N$-band ( $N\geq 2$) model, the
Berry curvature $\mathcal{F}_{\mu \nu }^{m}$ and the determinant of quantum
metric $g_{\mu \nu }^{m}$ of the $m$-th energy band satisfy the following
inequality: 
\begin{equation}
\det g_{\mu \nu }^{m}-\left( \frac{\mathcal{F}_{\mu {\nu }}^{m}}{2}\right)
^{2}\geq 0.
\end{equation}%
The equality holds if and only if $\left[ \mathbb{I}-\mathcal{P}\left( 
\mathbf{k}\right) \right] \left\vert {\partial _{{\nu }}u}_{m}\left( \mathbf{%
k}\right) \right\rangle $ differs from $\left[ \mathbb{I}-\mathcal{P}\left( 
\mathbf{k}\right) \right] \left\vert {\partial _{\mu }u}_{m}\left( \mathbf{k}%
\right) \right\rangle $ only a complex number. Especially, for $\forall $ $%
\mathbf{k}\in \mathcal{M}_{\mathbf{k}}$, if $N=2$ then we have the equality $%
\det g_{\mu \nu }^{m}-\left( \mathcal{F}_{\mu {\nu }}^{m}/2\right) ^{2}=0$.

\emph{Proof.}---\ Let us consider the quantum geometric tensor $Q_{\mu {\nu }%
}$ on the Bloch state manifold for the $m$-th energy band~\cite{Ma_PRB} 
\begin{eqnarray}
Q_{\mu {\nu }}^{m} &=&\langle {\partial _{\mu }u}_{m}\left( \mathbf{k}%
\right) |\left[ \mathbb{I}\text{ }\boldsymbol{-}\text{ }\left\vert {u}%
_{m}\left( \mathbf{k}\right) \right\rangle \left\langle {u}_{m}\left( 
\mathbf{k}\right) \right\vert \right] \left\vert {\partial _{{\nu }}u}%
_{m}\left( \mathbf{k}\right) \right\rangle  \notag \\
&=&g_{\mu \nu }^{m}-\frac{i}{2}\mathcal{F}_{\mu \nu }^{m},  \label{QGT}
\end{eqnarray}%
where ${u}_{m}\left( \mathbf{k}\right) $ is the period part of the Bloch
state for the $m$-th energy band, $\mu ,\nu =1,2$ denote the components of
the quasi-momentum $\mathbf{k}$, $g_{\mu \nu }^{m}=\Re Q_{\mu {\nu }}^{m}$
and $\mathcal{F}_{\mu \nu }^{m}=$ $-2\Im Q_{\mu {\nu }}^{m}$ are the quantum
metric and the Berry curvature of the $m$-th energy band, respectively. We
have%
\begin{eqnarray}
\det Q_{\mu {\nu }}^{m} &=&\left\vert 
\begin{array}{cc}
g_{\mu \mu }^{m} & g_{\mu {\nu }}^{m}-i\mathcal{F}_{\mu {\nu }}^{m}/2 \\ 
g_{\mu {\nu }}^{m}+i\mathcal{F}_{\mu {\nu }}^{m}/2 & g_{{\nu \nu }}^{m}%
\end{array}%
\right\vert  \notag \\
&=&\det g_{\mu {\nu }}^{m}-\left( \frac{\mathcal{F}_{\mu {\nu }}^{m}}{2}%
\right) ^{2}.  \label{detq}
\end{eqnarray}%
On the other hand, substituting the first line of the Eq. (\ref{QGT}) in Eq.
(\ref{detq}), and using the orthocomplement projection operator $\mathcal{P}%
^{\prime }:=\mathbb{I}-\mathcal{P}=\sum_{n\prime \neq m}\left\vert {u}%
_{n^{\prime }}\left( \mathbf{k}\right) \right\rangle \left\langle {u}%
_{n^{\prime }}\left( \mathbf{k}\right) \right\vert $, we can rewrite Eq. (%
\ref{detq}) as 
\begin{eqnarray}
&&\det Q_{\mu {\nu }}^{m}  \notag \\
&=&\left\vert 
\begin{array}{cc}
\langle {\partial _{\mu }u}_{m}\left( \mathbf{k}\right) |\mathcal{P}^{\prime
}\left\vert {\partial _{\mu }u}_{m}\left( \mathbf{k}\right) \right\rangle & 
\langle {\partial _{\mu }u}_{m}\left( \mathbf{k}\right) |\mathcal{P}^{\prime
}\left\vert {\partial _{{\nu }}u}_{m}\left( \mathbf{k}\right) \right\rangle
\\ 
\langle {\partial _{{\nu }}u}_{m}\left( \mathbf{k}\right) |\mathcal{P}%
^{\prime }\left\vert {\partial _{\mu }u}_{m}\left( \mathbf{k}\right)
\right\rangle & \langle {\partial _{{\nu }}u}_{m}\left( \mathbf{k}\right) |%
\mathcal{P}^{\prime }\left\vert {\partial _{{\nu }}u}_{m}\left( \mathbf{k}%
\right) \right\rangle%
\end{array}%
\right\vert  \notag \\
&=&\left\Vert \mathcal{P}^{\prime }\left\vert {\partial _{\mu }u}_{m}\left( 
\mathbf{k}\right) \right\rangle \right\Vert ^{2}\left\Vert \mathcal{P}%
^{\prime }\left\vert {\partial _{{\nu }}u}_{m}\left( \mathbf{k}\right)
\right\rangle \right\Vert ^{2}  \notag \\
&&-\left\Vert \langle {\partial _{\mu }u}_{m}\left( \mathbf{k}\right) |%
\mathcal{P}^{\prime }\cdot \mathcal{P}^{\prime }\left\vert {\partial _{{\nu }%
}u}_{m}\left( \mathbf{k}\right) \right\rangle \right\Vert ^{2}\geq 0.
\label{cauchy}
\end{eqnarray}%
In the last step, we made use of the Cauchy--Schwartz inequality and the
property of projection operator $\mathcal{P}^{\prime 2}=\mathcal{P}^{\prime
} $. Thus we prove that the relation $\det g_{\mu \nu }^{m}-\left( \mathcal{F%
}_{\mu {\nu }}^{m}/2\right) ^{2}\geq 0$ holds in a two-dimensional $N$-band
( $N\geq 2$) model. It is immediately clear that the equality holds if and
only if $\mathcal{P}^{\prime }\left\vert {\partial _{\mu }u}_{m}\left( 
\mathbf{k}\right) \right\rangle $ is parallel to $\mathcal{P}^{\prime
}\left\vert {\partial _{{\nu }}u}_{m}\left( \mathbf{k}\right) \right\rangle $%
, that is, there exists $c\in 
\mathbb{C}
$, such that 
$\left[ \mathbb{I}\text{ }\boldsymbol{-}\text{ }\left\vert {u}_{m}\left( 
\mathbf{k}\right) \right\rangle \left\langle {u}_{m}\left( \mathbf{k}\right)
\right\vert \right] \left\vert {\partial _{\mu }u}_{m}\left( \mathbf{k}%
\right) \right\rangle =c$ $\left[ \mathbb{I}\text{ }\boldsymbol{-}\text{ }%
\left\vert {u}_{m}\left( \mathbf{k}\right) \right\rangle \left\langle {u}%
_{m}\left( \mathbf{k}\right) \right\vert \right] \left\vert {\partial _{{\nu 
}}u}_{m}\left( \mathbf{k}\right) \right\rangle .$ 

In the two-band case, we assume the eigenstates of the Hamiltonian are $%
\left\{ \left\vert {u}_{-}\left( \mathbf{k}\right) \right\rangle ,\left\vert 
{u}_{+}\left( \mathbf{k}\right) \right\rangle \right\} $, and then, we
define $\mathcal{P}\left( \mathbf{k}\right) =\left\vert {u}_{-}\left( 
\mathbf{k}\right) \right\rangle \left\langle {u}_{-}\left( \mathbf{k}\right)
\right\vert \ $and $\mathcal{P}^{\prime }\left( \mathbf{k}\right)
=\left\vert {u}_{+}\left( \mathbf{k}\right) \right\rangle \left\langle {u}%
_{+}\left( \mathbf{k}\right) \right\vert $. For $\forall $ $\mathbf{k}\in 
\mathcal{M}_{\mathbf{k}}$, it is easy to see that $\mathcal{P}^{\prime
}\left( \mathbf{k}\right) \left\vert {\partial _{\mu }u}_{-}\left( \mathbf{k}%
\right) \right\rangle \ $is parallel to $\mathcal{P}^{\prime }\left( \mathbf{%
k}\right) \left\vert {\partial _{{\nu }}u}_{-}\left( \mathbf{k}\right)
\right\rangle .$ This means that, in a two-dimensional two-band system, the
relation $\det g_{\mu \nu }^{m}=\left( \mathcal{F}_{\mu {\nu }}^{m}/2\right)
^{2}$ always holds. The proof is completed. $\hfill \blacksquare $

\subsection{Two-band system: what conditions make the quantum metric =
Riemann metric}

\emph{Theorem 2.}--- In a general two-dimensional two-band system, the
positive semi-definite quantum metric $g_{\mu {\nu }}$ to be a proper
Riemann metric if and only if the Berry curvature $\mathcal{F}_{\mu \nu
}\neq 0$ in the first Brillouin zone.

\emph{Proof.}--- In consideration of the positive semi-definite of quantum
metric $g_{\mu {\nu }}$ and the Theorem 1, for the $m$-th energy band, if
and only if $\mathcal{F}_{\mu \nu }\neq 0,$ we can obtain%
\begin{equation}
\det g_{\mu \nu }^{m}=\left( \frac{\mathcal{F}_{\mu {\nu }}^{m}}{2}\right)
^{2}>0.
\end{equation}%
The proof is completed. $\hfill \blacksquare $

This theorem shows that, in two-dimensional two-band system, if exists a
non-vanishing Berry curvature $\mathcal{F}_{\mu {\nu }}^{m}$ in the entire
first Brillouin zone, then the quantum metric $g_{\mu \nu }^{m}$ defines a
proper Riemann metric on the Bloch band manifold.

\section{Some examples}

{\label{example} }

\subsection{A time-reversal-symmetric two-band model}

This model has been discussed in Ref.~\cite{Ma2013} and Ref.~\cite{xtan},
the Bloch Hamiltonian reads $\mathcal{H}(k,\varphi )=\sum_{\alpha
=1}^{3}d_{\alpha }\left( k,\varphi \right) \sigma ^{\alpha }$, where $%
d_{1}\left( k,\varphi \right) =\frac{1}{2}\gamma \sin k\sin 2\varphi $, $%
d_{2}\left( k,\varphi \right) =\frac{1}{2}\gamma \sin k\cos 2\varphi $, $%
d_{3}\left( k,\varphi \right) =-\frac{1}{2}\left( h+\cos k\right) $ with $%
k\in \lbrack -\pi ,\pi ]$ and $\varphi \in \lbrack 0,\pi ]$, $h$, $\gamma $ (%
$\gamma \neq 0$) are the Hamiltonian parameters, and $\sigma ^{\alpha }$
denote the three Pauli matrices.\ The quantum metric and the Berry curvature
for the occupied lower band are given by%
\begin{equation}
g_{k\varphi }\left( h,\gamma \right) =\frac{1}{4}\left( 
\begin{array}{cc}
\frac{\gamma ^{2}\left( 1+h\cos k\right) ^{2}}{\left( h+\cos k\right)
^{2}+\gamma ^{2}\sin ^{2}k} & 0 \\ 
0 & \frac{2\gamma ^{2}\sin ^{2}k}{\left( h+\cos k\right) ^{2}+\gamma
^{2}\sin ^{2}k}%
\end{array}%
\right) ,  \label{g}
\end{equation}%
and%
\begin{equation}
\mathcal{F}_{k\varphi }\left( h,\gamma \right) =\frac{\gamma ^{2}\left(
1+h\cos k\right) \sin k}{\left[ \left( h+\cos k\right) ^{2}+\gamma ^{2}\sin
^{2}k\right] ^{3/2}},  \label{f}
\end{equation}%
respectively \cite{note}.

\begin{figure}[tbp]
\includegraphics[width=3in]{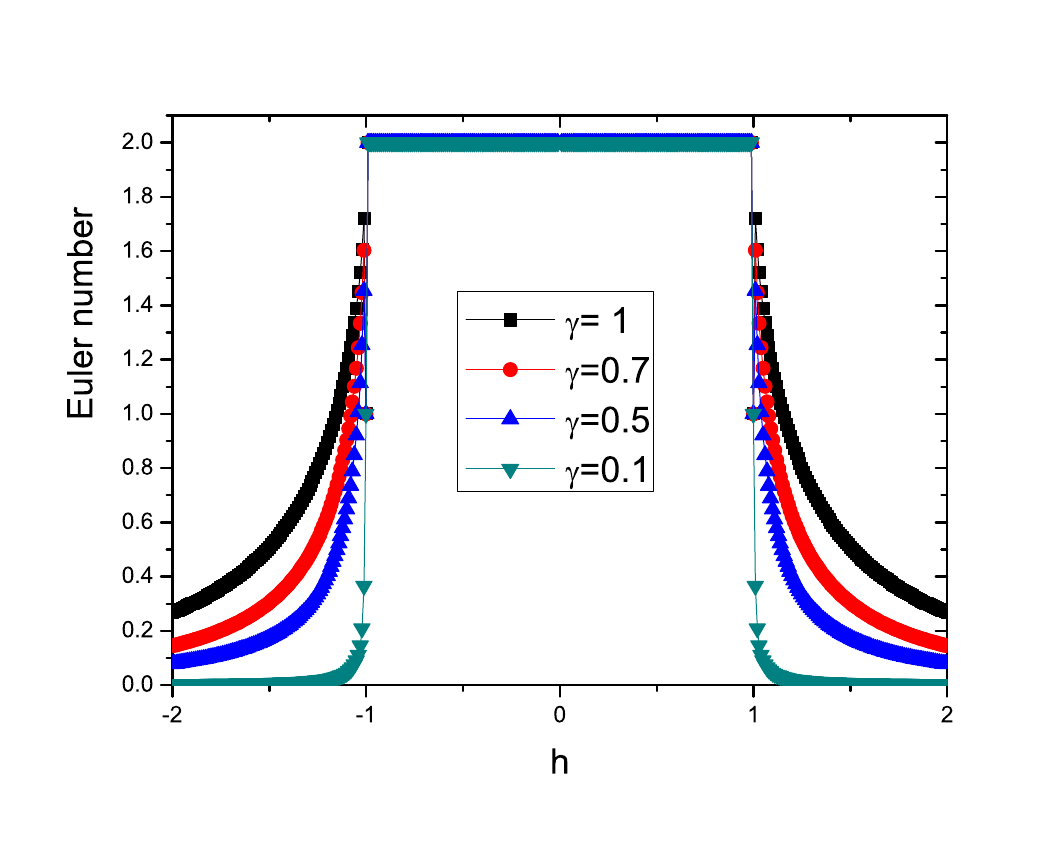}
\caption{(color online) The Euler number $\protect\chi =2$ in the
topological nontrivial phase $\left\vert h\right\vert <1$, and ill-defined
in the topological trivial phase $\left\vert h\right\vert >1$.}
\label{Euler}
\end{figure}

As shown in Ref.~\cite{Ma2013} and Ref.~\cite{xtan}, even though the Chern
number for the time-reversal-symmetric system is trivial, the system has a
nontrivial Euler number of the band in the topological nontrivial phase (see
Fig.~\ref{Euler}). Note that the occupied Bloch band forms a two-dimensional
closed manifold on the first Brillouin zone $T^{2}$, which equipped with the
quantum metric $g_{k\varphi }\left( h,\gamma \right) $ pulled back from the
Bloch line bundle. The Euler characteristic number can be calculated
conveniently by the Gauss-Bonnet theorem $\chi =\left( 2\pi \right)
^{-1}\int_{\text{Bz}}\mathcal{K}dA$, where $\mathcal{K}$ is the Gauss
curvature, $dA=\det^{1/2}g_{k\varphi }\left( h,\gamma \right) d{k}d\varphi $
denotes the Brillouin zone area element measured by the quantum metric.
Finally, we can derive the Euler number of the occupied Bloch band (for
details see Ref. \cite{Ma2013}) 
\begin{eqnarray}
\chi  &=&\frac{1}{2\pi }\int_{\text{Bz}}\mathcal{K}dA  \notag \\
&=&\frac{1}{2\pi }\iint_{\text{Bz}}\left\vert \frac{\vec{d}\cdot {\partial
_{k}}\vec{d}\times {\partial _{\varphi }}\vec{d}}{d^{3}}\right\vert d{k}%
d\varphi   \notag \\
&=&\left\{ 
\begin{array}{lll}
1 &  & \text{if}\quad \left\vert h\right\vert <1; \\ 
\text{Metric degenerate} &  & \text{if}\quad \left\vert h\right\vert >1,%
\text{ill defined. }%
\end{array}%
\right. .
\end{eqnarray}

Despite the isolated singular points ${k=0,\pm \pi }$ (which is independent
on the Hamiltonian parameter $h,\gamma $), it is easy to find from the Eq. (%
\ref{g}) that if the parameter $\left\vert h\right\vert >1$, then the
quantum metric must be degenerate in some region of the Brillouin zone where 
$1+hk=0$. According to the Theorem 2, it can be understood intuitively as
follows: if $\left\vert h\right\vert >1,$ then the monopole (anti-monopole)
is not enclosed by the surface $\vec{d}(k,\varphi )$. This means that the
Berry curvature as a solid angle to the monopole can continuously tend to
zero in some region of the Brillouin zone, and then leads to a degenerate
quantum metric $\det g_{k\varphi }\left( h,\gamma \right) =0$.

\begin{figure}[tbh]
\begin{center}
\includegraphics[width=0.23\textwidth,height=0.15\textheight]{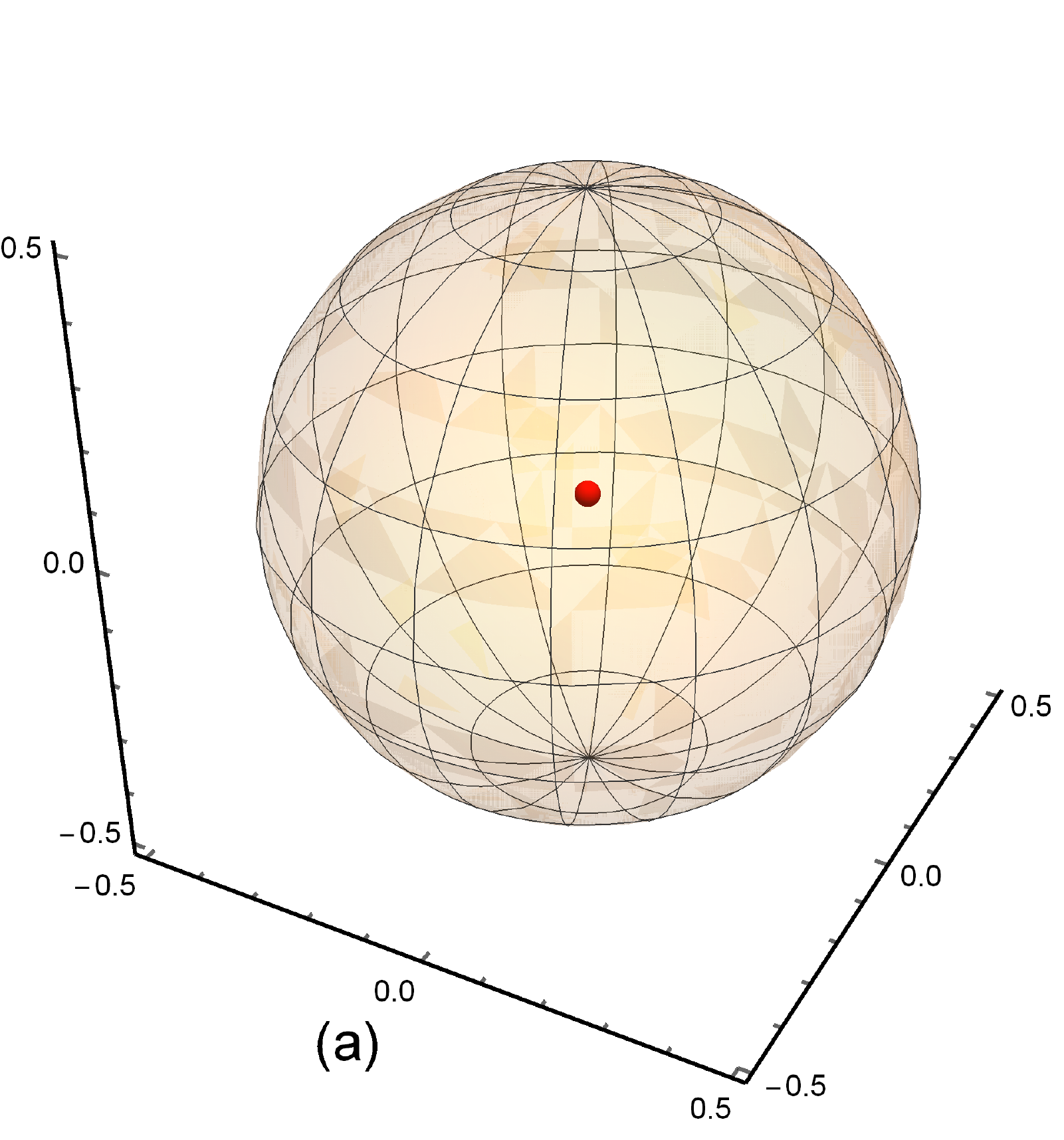} %
\includegraphics[width=0.23\textwidth,height=0.15\textheight]{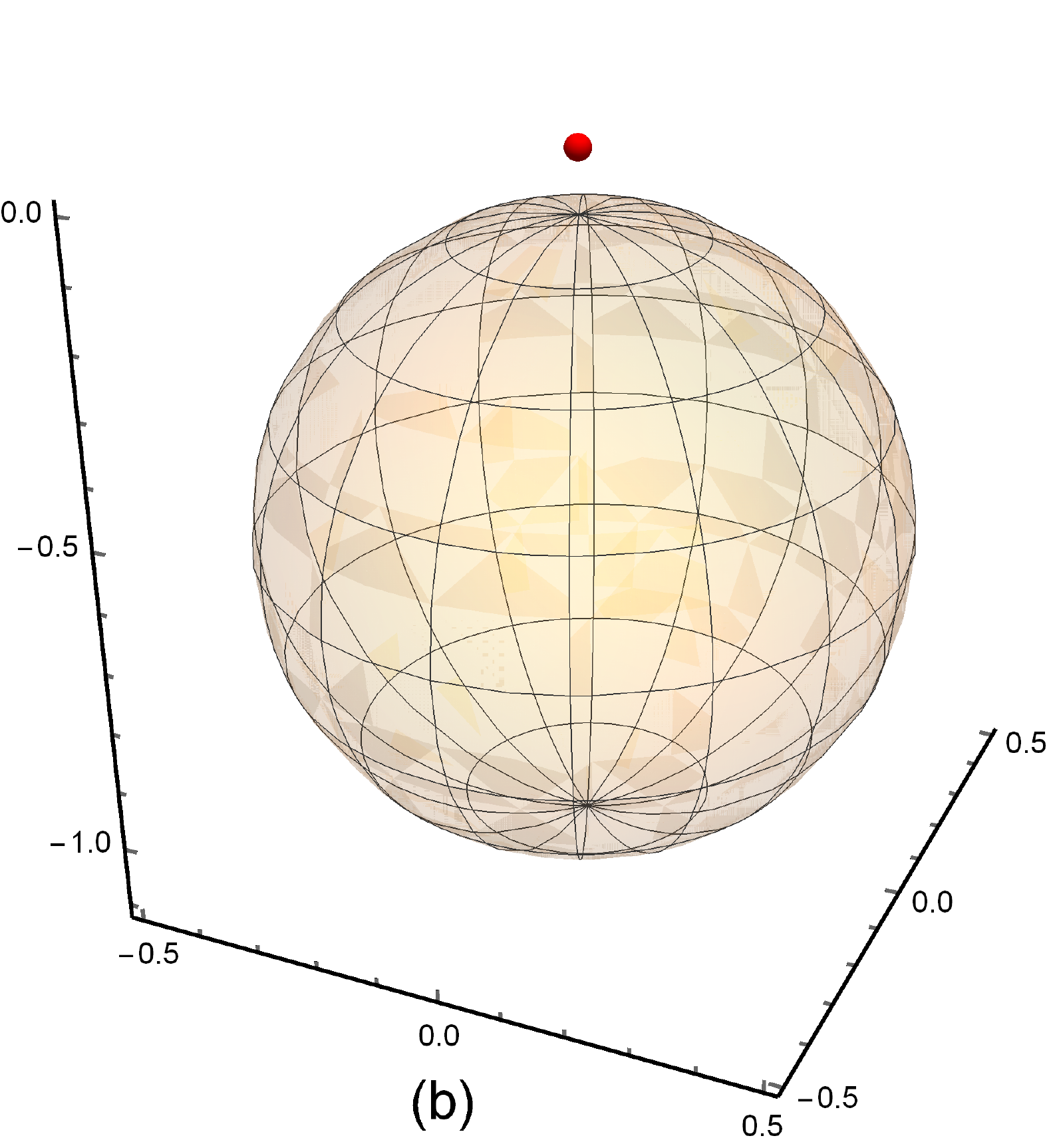} %
\includegraphics[width=0.23\textwidth,height=0.15\textheight]{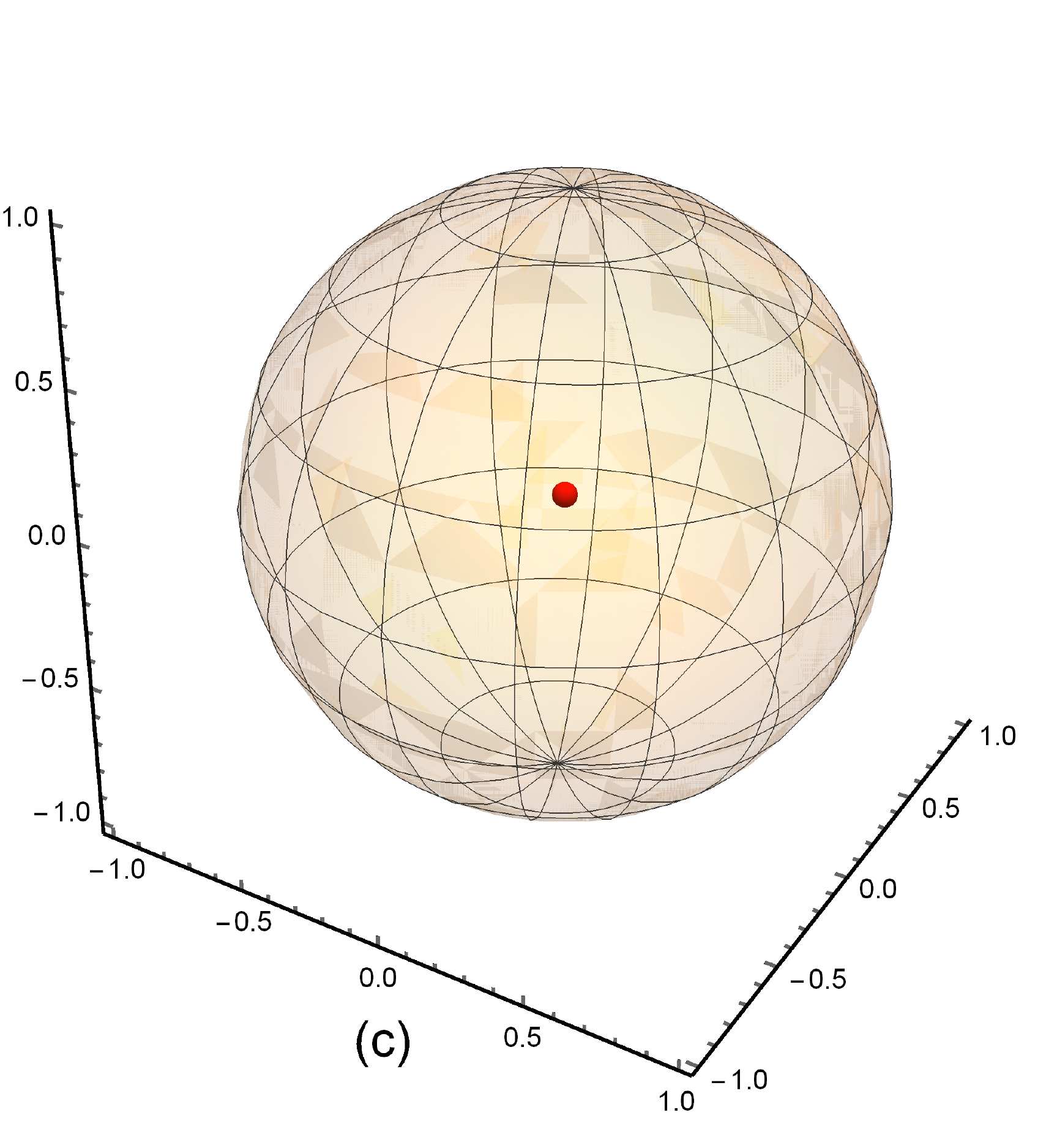} %
\includegraphics[width=0.23\textwidth,height=0.15\textheight]{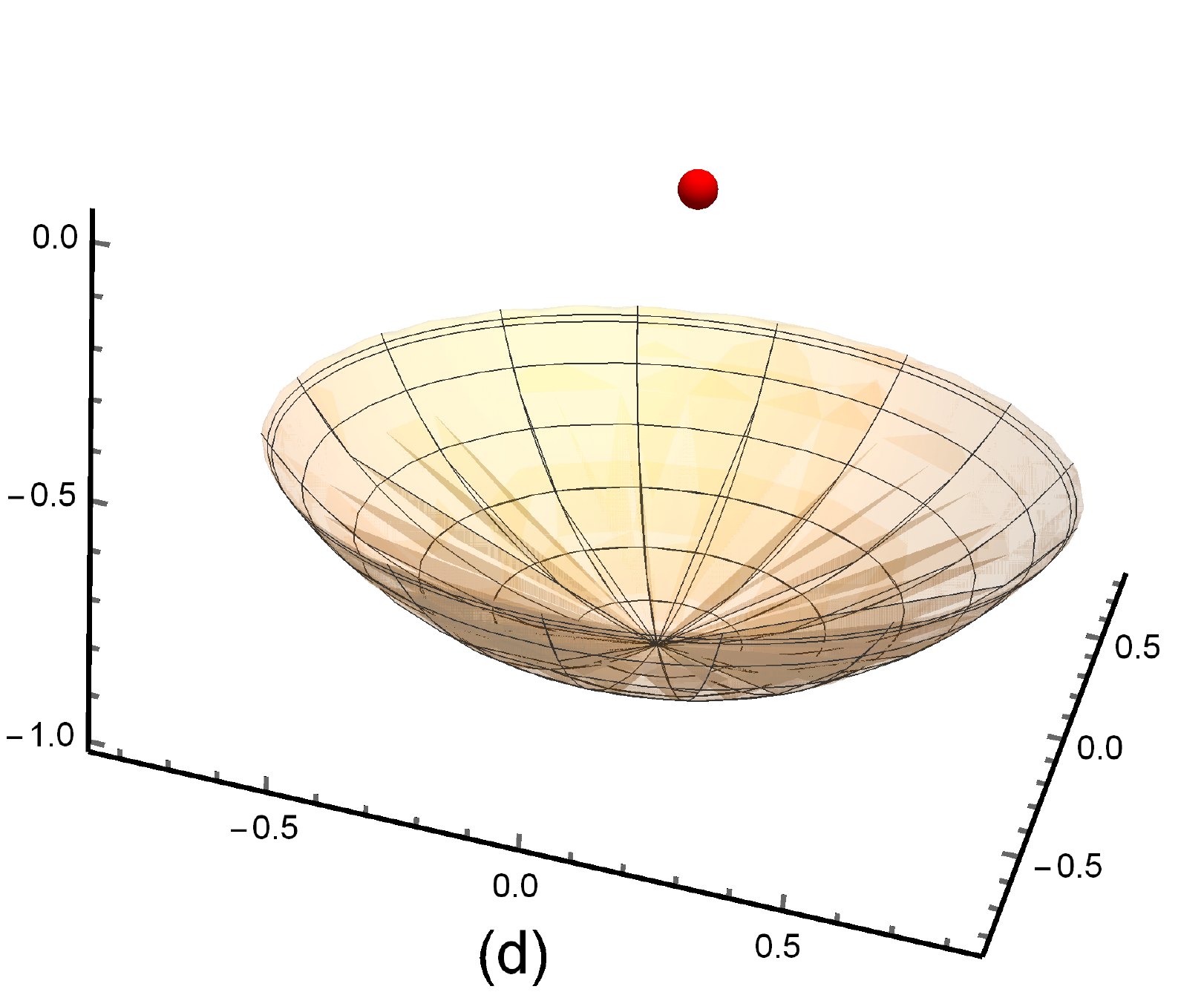}
\end{center}
\caption{(color online) The trajectories of the vector $\vec{d}\left( k,%
\protect\varphi \right) $ with the Hamiltonian parameters: (a) $h=0,\protect%
\gamma =1$, (b) $h=1.2,\protect\gamma =1$; The trajectories of the unit
vector $\vec{d}\left( k,\protect\varphi \right) /d\left( k,\protect\varphi %
\right) $ with the Hamiltonian parameters: (c) $h=0,\protect\gamma =1$, (d) $%
h=1.2,\protect\gamma =1$. The position of the monopole (coordinates origin)
is marked by a red ball.}
\label{xychain}
\end{figure}

Finally, we illustrate why the surface of the unit vector $\vec{d}(k,\varphi
)/d(k,\varphi )$ is unclosed when the parameter $\left\vert h\right\vert >1$%
. The reason is straightforward that: if the parameter $\left\vert
h\right\vert >1$, then the monopole (coordinate origin) will be not enclosed
by the surface $\vec{d}(k,\varphi )$ (see Fig.~\ref{xychain}$b$). Hence, if
we use the unit vector $\vec{d}(k,\varphi )/d(k,\varphi )$, then the
normalized operation $\vec{d}(k,\varphi )/d(k,\varphi )$ will squashed the
two surfaces (on the same side of the monopole) flat and make it exhibit a
boundary (see Fig.~\ref{xychain}$d$).

\subsection{Haldane model}
{\label{example b} }

The Haldane model Hamiltonian can be expressed in the two-component
representation $\left( c_{\mathbf{k}A}^{\dagger },c_{\mathbf{k}B}^{\dagger
}\right) $ associated with sublattices A and B as $\mathcal{H}(\mathbf{k}%
)=\epsilon \left( \mathbf{k}\right) \mathbb{I}+\sum_{\alpha =1}^{3}d_{\alpha
}\left( \mathbf{k}\right) \sigma ^{\alpha }$. Choosing the proper reciprocal
lattice vectors, we have 
\begin{eqnarray}
\epsilon \left( \mathbf{k}\right) &=&2t_{2}\cos \varphi \left[ \cos
(k_{x})+\cos (k_{y})-\cos (k_{x}+k_{y})\right] ,  \notag \\
d_{1}\left( \mathbf{k}\right) &=&t_{1}\left[ 
\begin{array}{c}
\cos \left( 2k_{x}/3+k_{y}/3\right) +\cos \left( k_{x}/3+2k_{y}/3\right) \\ 
+\cos \left( k_{x}/3-k_{y}/3\right)%
\end{array}%
\right] ,  \notag \\
d_{2}\left( \mathbf{k}\right) &=&t_{1}\left[ 
\begin{array}{c}
\sin \left( 2k_{x}/3+k_{y}/3\right) -\sin \left( k_{x}/3+2k_{y}/3\right) \\ 
-\sin \left( k_{x}/3-k_{y}/3\right)%
\end{array}%
\right] ,  \notag \\
d_{3}\left( \mathbf{k}\right) &=&t_{0}-2t_{2}\sin \varphi \left[ 
\begin{array}{c}
\sin (k_{x})+\sin (k_{y}) \\ 
-\sin (k_{x}+k_{y})%
\end{array}%
\right] .
\end{eqnarray}%
According to the Theorem 2, we can use the Berry curvature of the energy
band as a tool to determine whether the metric is a proper Riemann metric~%
\cite{note2}.

\begin{figure}[tbh]
\includegraphics[width=2.6in]{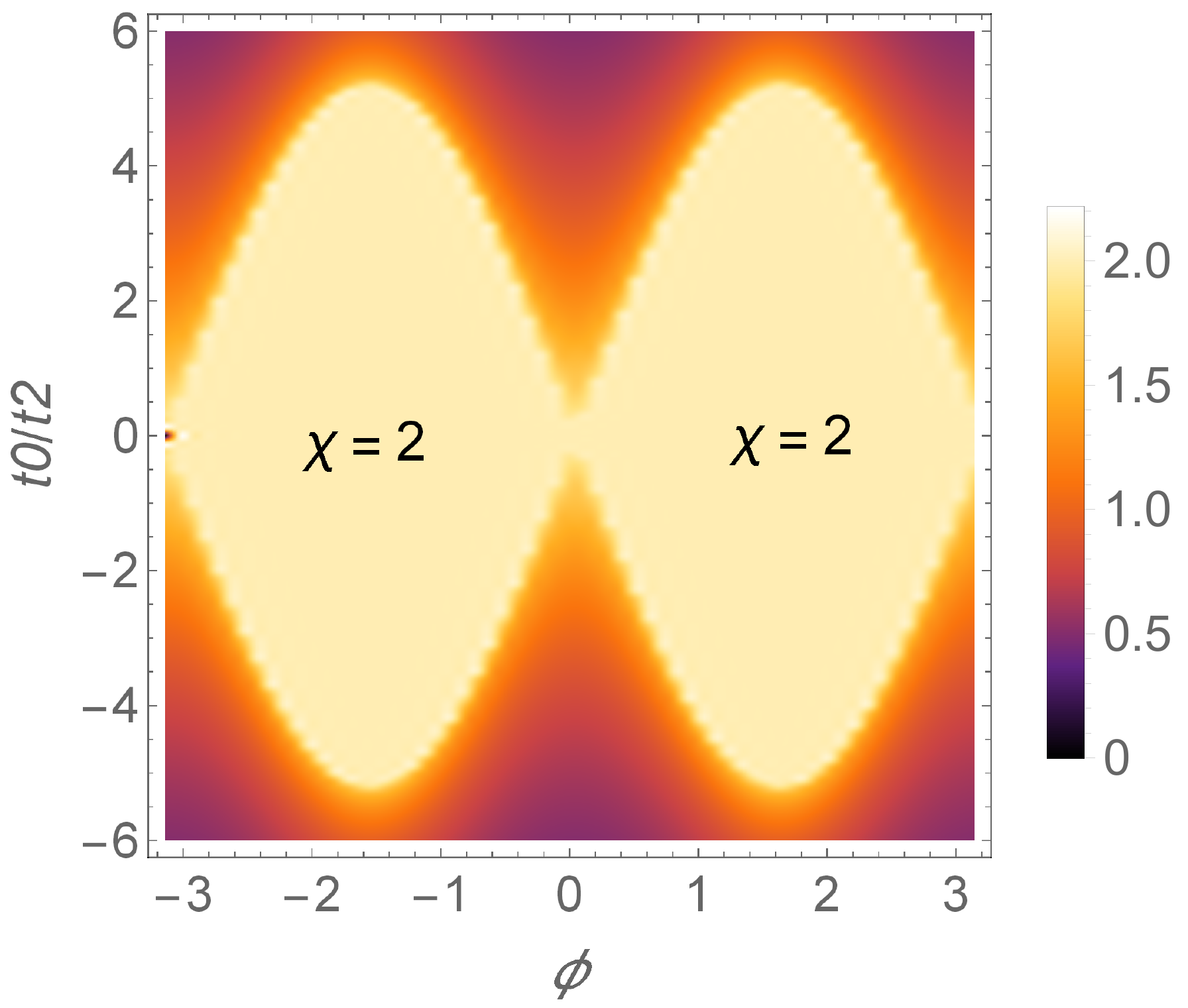}
\caption{(color online) The phase diagram of Haldane model marked by Euler
number of the energy band. The Euler number $\protect\chi =2$ in the
topological nontrivial phase $C_{1}=\pm 1$, and is ill-defined in the
topological trivial phase $C_{1}=0,$ where the quantum metric is degenerate
and its inverse does not exist.}
\label{Euler_H}
\end{figure}

\begin{figure}[tbh]
\begin{center}
\includegraphics[width=0.23\textwidth,height=0.15\textheight]{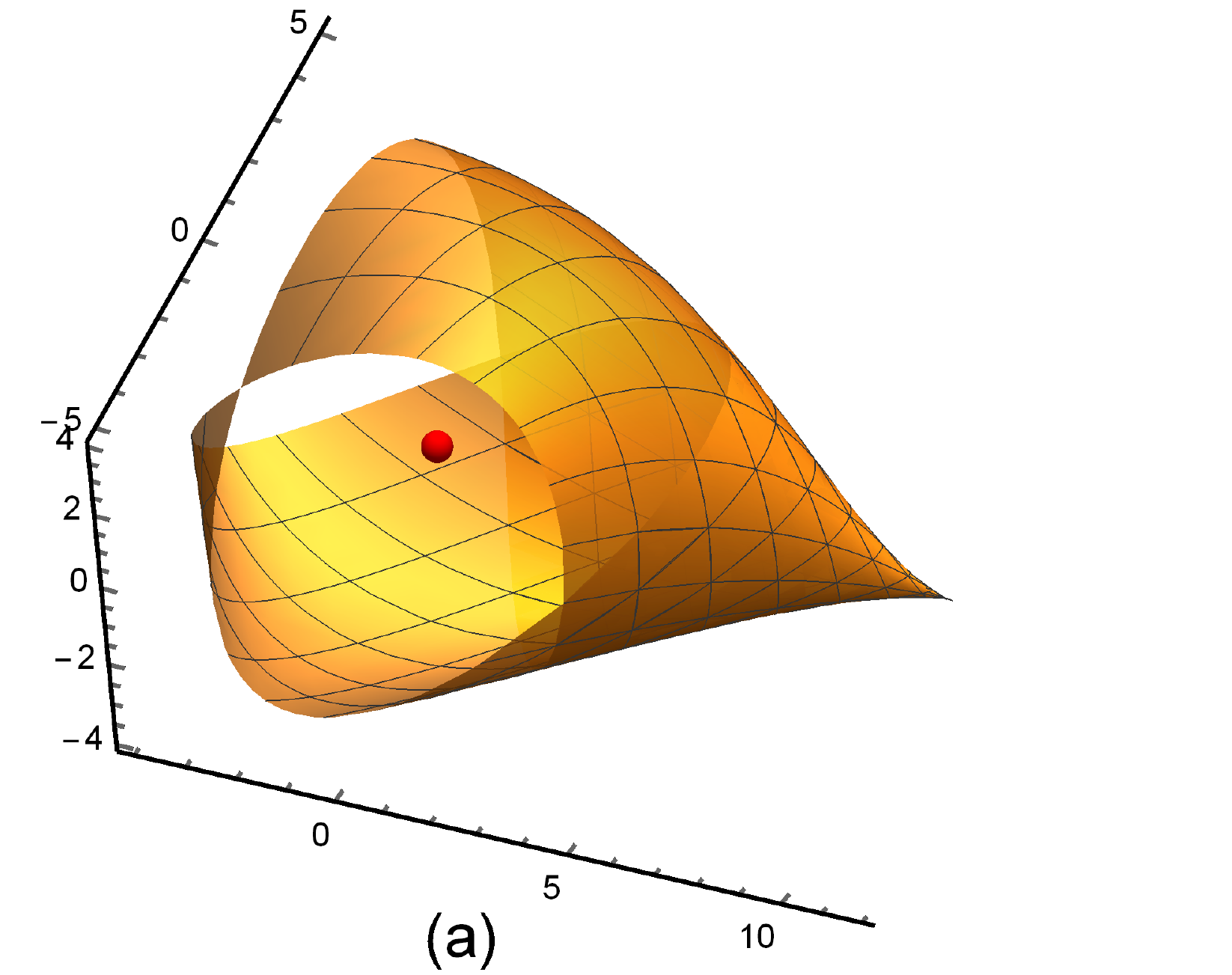} %
\includegraphics[width=0.23\textwidth,height=0.15\textheight]{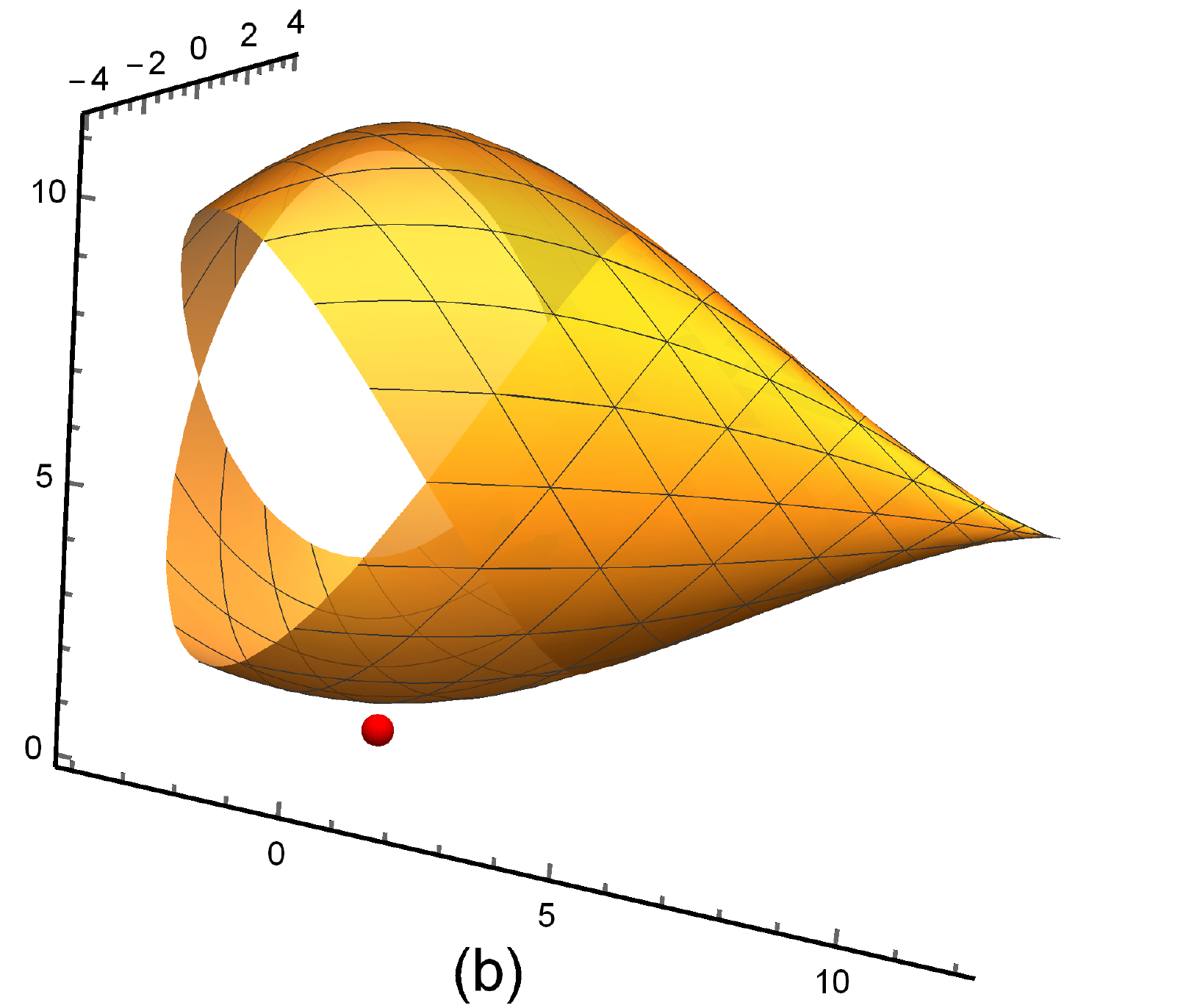} %
\includegraphics[width=0.23\textwidth,height=0.15\textheight]{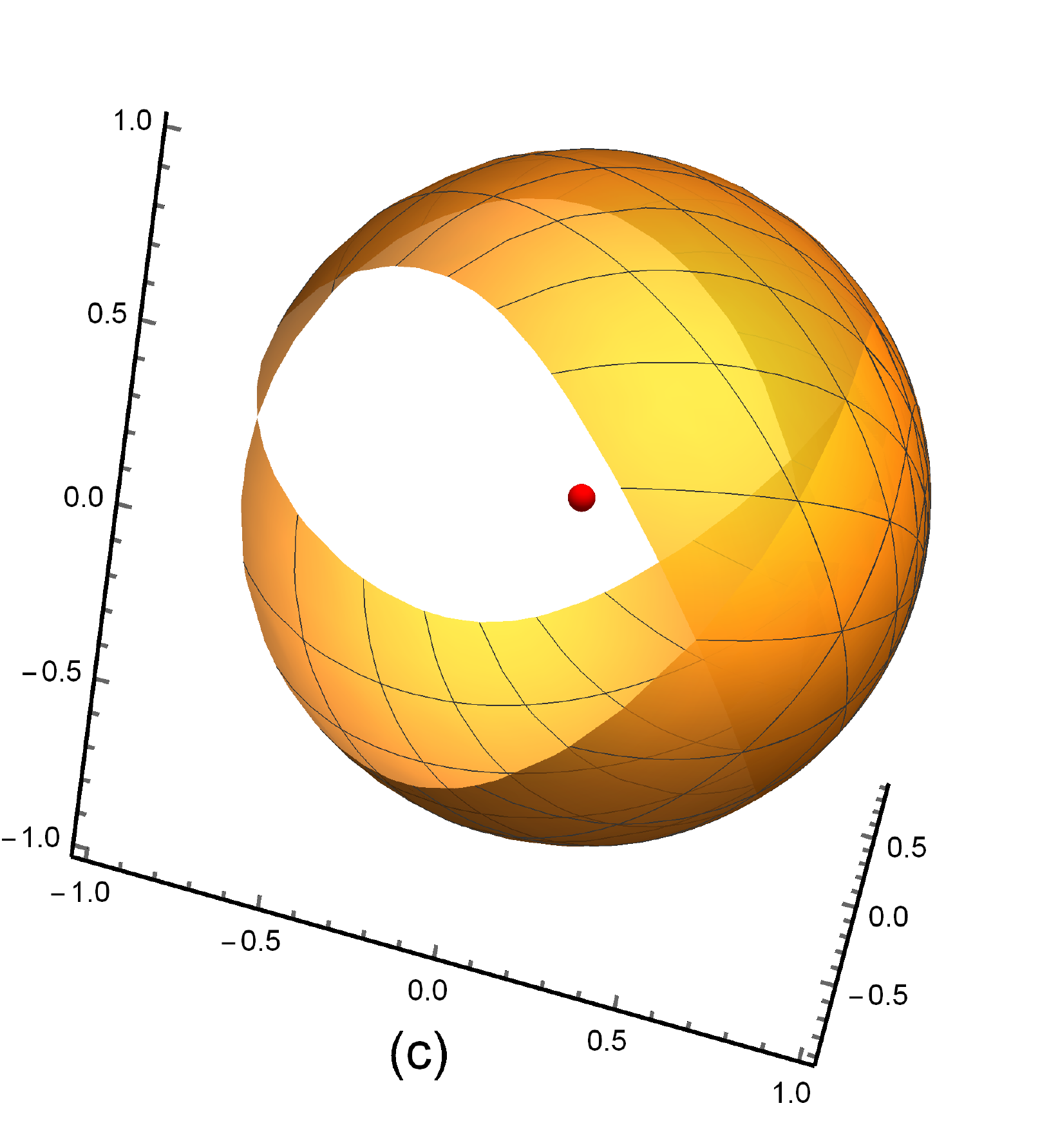} %
\includegraphics[width=0.23\textwidth,height=0.15\textheight]{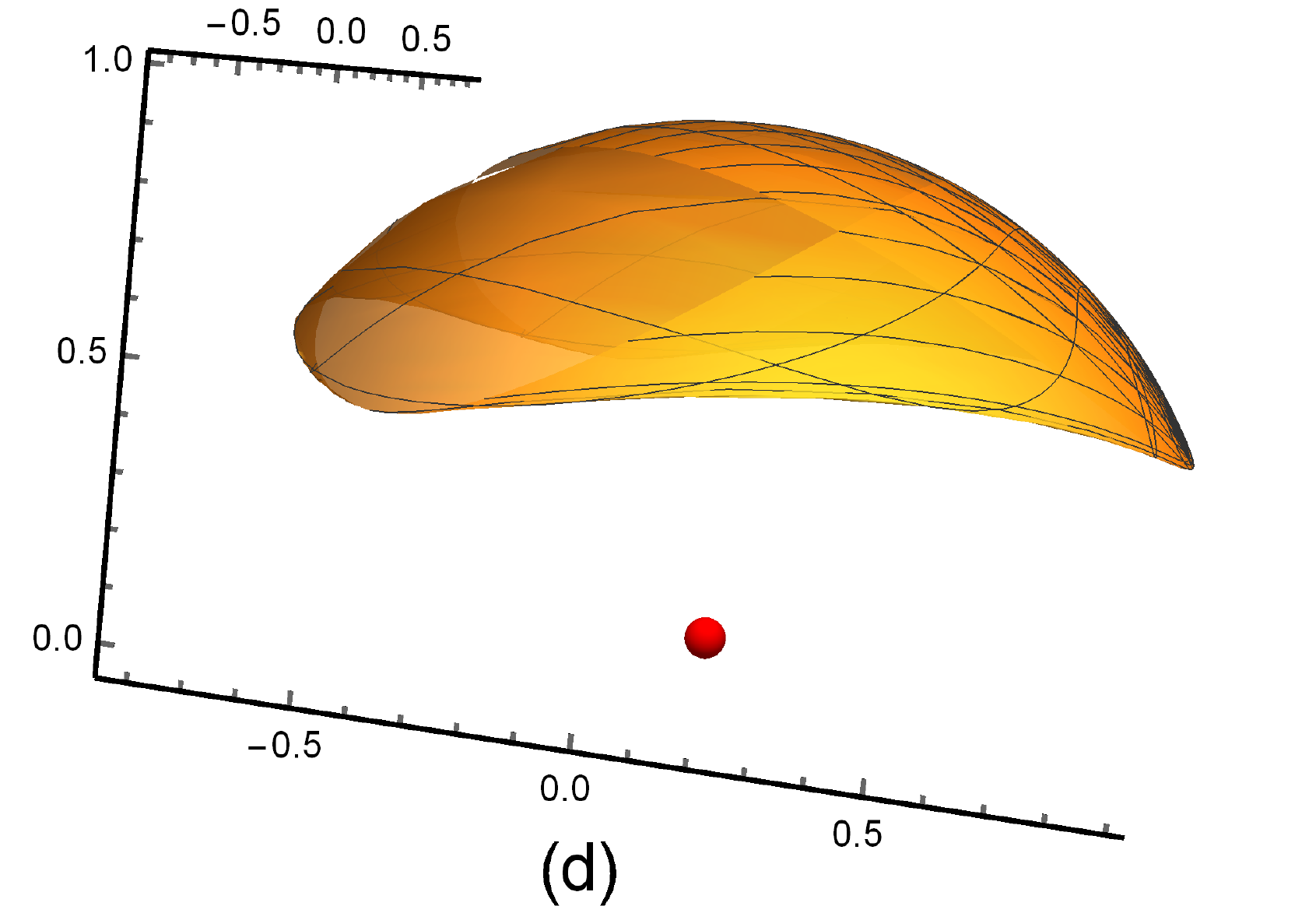}
\end{center}
\caption{(color online) The trajectories of the vector $\vec{d}\left( 
\mathbf{k}\right) $ with the Hamiltonian parameters: (a) $%
t_{0}=0,t_{1}=4,t_{2}=1,\protect\varphi =\protect\pi /2;$ (b) $t_{0}=3%
\protect\sqrt{3+1},t_{1}=4,t_{2}=1,\protect\varphi =\protect\pi /2;$ The
trajectories of the unit vector $\vec{d}\left( \mathbf{k}\right) /d\left( 
\mathbf{k}\right) $ with the Hamiltonian parameters: (c) $%
t_{0}=0,t_{1}=4,t_{2}=1,\protect\varphi =\protect\pi /2;$ (d) $t_{0}=3%
\protect\sqrt{3+1},t_{1}=4,t_{2}=1,\protect\varphi =\protect\pi /2.$ The
position of the monopole (coordinates origin) is marked by a red ball.}
\label{haldane}
\end{figure}

As shown in the Fig.~\ref{Euler_H}, we find that the Euler number of the
filled band is $\chi =2$ corresponding to the Chern number $C_{1}=\pm 1$,
and $\chi =$ \textquotedblleft non-integer\textquotedblright\ corresponding
to the $C_{1}=0$. It is worth pointing out that the surface of $\vec{d}(%
\mathbf{k})/d(\mathbf{k})$ for the Haldane Hamiltonian are all unclosed no
matter in the topological nontrivial phase or in the trivial phase. It can
be seen that, in the topological nontrivial phase, the Euler number for the
surface $\vec{d}(\mathbf{k})/d(\mathbf{k})$ itself should be zero due to two
boundaries existing on its surface (see Fig.~\ref{haldane}$a$ or \ref%
{haldane}$c$), which makes it topological equivalent to the $S^{1}$ with the
Euler number $\chi =0$. However, this is contradictory to the Euler number $%
\chi =2$ of the occupied band through the Gauss-Bonnet theorem in the
topological nontrivial phase. This shows that there is no correspondence
between the Euler number of the Bloch band and the Euler number of the $\vec{%
d}(\mathbf{k})/d(\mathbf{k})$ surface itself.

In addition, as shown in Fig.~\ref{haldane}$d$, it is not surprised to find
that the normalized operation $\vec{d}(\mathbf{k})/d(\mathbf{k})$ will
squashed the two surfaces (on the same side of the monopole) flat and make
it exhibit a boundary. The role of the mapping $\vec{d}(\mathbf{k})/d(%
\mathbf{k})$ unit sphere $S^{2}$ is actually to serve as a states space $%
\mathcal{C}P^{1}$ for the Fubini-Study metric, but not a parameter manifold
for it. However, the quantum metric is defined on a parameter manifold, and
the former is only a special case of the quantum metric.

\section{Conclusions}

In summary, we point out that a well-defined Euler number of the energy band
requires a positive definite quantum metric over the first Brillouin zone.
We also provide a sufficient criterion for a positive definite quantum
metric in a two-dimensional multi-band system based on the corresponding
Berry curvature. Furthermore, we show that this criterion is sufficient and
necessary in the two-band case. On the other hand, it should be noted that a
topological phase characterized by a nontrivial Chern number can not yet
ensure the Berry curvature is non-vanishing in the whole Brillouin zone, and
hence, a well-defined Euler number of the energy band requires more stronger
conditions than a nontrivial Chern number.

It is also interesting to note that the Euler number brings a topological
explanation to the notion of the band's \textquotedblleft
complexity\textquotedblright, which is introduced by Marzari and Vanderbilt
to measure the variation of the band projection operator throughout the
Brillouin zone~\cite{MV}. The \textquotedblleft complexity\textquotedblright\ is
defined by a gauge invariant Brillouin zone area measured by the quantum
metric $\int_{\text{Bz}}d\mathbf{k}\det^{1/2}g\left( \mathbf{k}\right) $. As
we now know, this quantity may be a topological number, which differs from
the Euler number of the band $\chi =\left( 2\pi \right) ^{-1}\int_{\text{Bz}}%
\mathcal{K}dA$ only a constant coefficient if the Berry curvature is
non-vanishing in the whole two-dimensional Brillouin zone.

\section{Acknowledgments}

The author thanks Shi-Liang Zhu and Peng Liu for helpful discussions. This
work was supported by the NSF of Beijing under Grant No. 1173011, the
Scientific Research Project of Beijing Municipal Education Commission (BMEC)
under Grant No. KM201711232019, and the Qin Xin Talents Cultivation Program
of Beijing Information Science and Technology University (BISTU) under Grant
No. QXTCP C201711.


\begin{thebibliography}{9}
\bibitem{note} Here we can see $g_{k\varphi }\left( h,\gamma \right) $ and $%
\mathcal{F}_{k\varphi }\left( h,\gamma \right) $ are all independent on the $%
\varphi $, and this can be explained by the following reasons: This model is
derived from a one-dimensional BDI class model $\mathcal{H}_{\text{1D}%
}(k)=\sum_{\alpha =1}^{3}d_{\alpha }\left( k\right) \sigma ^{\alpha }$, with 
$d_{1}\left( k\right) =0$, $d_{2}\left( k\right) =\frac{1}{2}\gamma \sin k$, 
$d_{3}\left( k\right) =-\frac{1}{2}\left( h+\cos k\right) $, by the
following $U\left( 1\right) $ gauge transformation $\mathcal{H}(k,\varphi )=$
$g(\varphi )\mathcal{H}_{\text{1D}}(k)g(\varphi )^{\dag }$ with $g(\varphi
)=\prod_{j=1}^{l-1}e^{i\varphi \left( t\right) c_{j}^{\dagger }c_{j}}$.
Obviously, both $g_{k\varphi }\left( h,\gamma \right) $ and $\mathcal{F}%
_{k\varphi }\left( h,\gamma \right) $ must be independent on the parameter $%
\varphi $ owing to their $U(1)$ gauge invariance. On the other hand, we have 
$\mathcal{F}_{k\varphi }\left( h,\gamma \right) =-\mathcal{F}_{-k-\varphi
}\left( h,\gamma \right) $ due to the time-reversal symmetry of $\mathcal{H}%
(k,\varphi )$, which leads to a vanishing Chern number and fails to
characterize the topological charge by using Chern number of the band.
\bibitem{note2} The Berry curvature of the lower occupied band in the
Haldane model is given by $\mathcal{F}_{xy}\left( t_{0},t_{1},t_{2},\varphi
\right) =\frac{t_{1}^{2}}{6}\left( 2t_{2}\sin \varphi \frac{B}{A}-t_{0}\frac{%
C}{A}\right) $, where 
\begin{eqnarray*}
A &=&\left[ 
\begin{array}{c}
\left[ t_{0}-2t_{2}\sin \varphi \left( \sin x-\sin (x+y)+\sin y\right) %
\right] ^{2} \\ 
+t_{1}^{2}\left( 2\cos x+2\cos (x+y)+2\cos y+3\right)%
\end{array}%
\right] ^{3/2}, \\
B &=&(\cos x-1)^{2}+(\cos y-1)^{2}+\left[ \cos (x+y)-1\right] ^{2}, \\
C &=&\sin x-\sin (x+y)+\sin y.
\end{eqnarray*}
\end{thebibliography}
\end{document}